\documentclass[11pt,a4paper]{article}
\pdfoutput=1
\usepackage{jheppub}
\usepackage{amsmath,amscd}
\usepackage{amsfonts}
\usepackage{amssymb}
\usepackage{graphicx}
\usepackage{color}
\usepackage[normalem]{ulem}
\usepackage{hyperref}
\usepackage{mathtools}

\newcommand{\CC}{\mathbb{C}} 
\newcommand{\RR}{\mathbb{R}} 
\newcommand{\ZZ}{\mathbb{Z}} 

\def\tr         {{\rm  tr}}

\def\calc         {{\cal C}}

\def\be{\begin{equation}}
\def\ee{\end{equation}}
\def\bea{\begin{eqnarray}}
\def\eea{\end{eqnarray}}

\def\b{\beta}

\def\g{\gamma}

\def\l{\lambda}

\def\f{\phi}

\def\O{\Omega}
\def\p{\pi}

\def\s{\sigma}

\def\t{\tau}

\def\sF{{{ F}\!\!\!\!\hskip.8pt\hbox{\raise1pt\hbox{/}}\,}}
\def\som{{{ \omega}\!\!\!\!\hskip.8pt\hbox{\raise1pt\hbox{/}}\,}}
\def\sJ{{{\rm J}\!\!\!\!\hskip.8pt\hbox{\raise1pt\hbox{/}}\,}}

\def\pa{\partial}

\def\to{\rightarrow}
\def\nonu{\nonumber \\{}}
\def\half{{1 \over 2}}

\newcommand{\commout}[1]{}

\title{A note on the admissibility of complex BTZ metrics}

\author[a]{Ivano Basile,}
\author[b,1]{Andrea Campoleoni\note{Research Associate of the Fund for Scientific Research -- FNRS, Belgium.}}
\author[c]{and Joris Raeymaekers}

\affiliation[a]{Arnold Sommerfeld Center for Theoretical Physics, Ludwig Maximilians Universit\"at M\"unchen, Theresienstrasse 37, 80333 M\"unchen, Germany.}
\emailAdd{ivano.basile@lmu.de}
\affiliation[b]{Service de Physique de l'Univers, Champs et Gravitation, Universit\'{e} de Mons -- UMONS, \\ Place du Parc 20, 7000 Mons, Belgium}
\emailAdd{andrea.campoleoni@umons.ac.be}
\affiliation[c]{CEICO, Institute of Physics of the ASCR,  Na Slovance 2, 182 21 Prague 8, Czech Republic.}
\emailAdd{joris@fzu.cz}

\abstract{We perform a nontrivial check of Witten's recently proposed admissibility criterion for complex metrics. We consider the `quasi-Euclidean' metrics obtained from continuing the BTZ class of metrics to imaginary time. Of special interest are the overspinning metrics, which are smooth in this three-dimensional context. Their inclusion as saddle points in the gravitational path integral would lead to puzzling results in conflict with those obtained using other methods. It is therefore encouraging that the admissibility criterion discards them. For completeness, we perform an analysis of smoothness and admissibility for the family of quasi-Euclidean BTZ metrics at all values of the mass and angular momentum.}
\arxivnumber{}
\keywords{}
\begin{document}

\maketitle
\section{Introduction and summary}
In the  path-integral approach to quantum gravity pioneered by Gibbons and Hawking \cite{Gibbons:1976ue}, it was clear early on that in some cases complex metrics should be allowed to contribute to the `Euclidean' path integral.  For example, the thermodynamical properties of rotating black holes follow from admitting a complex saddle point  in the path integral.

On the other hand, integrating over {\em all} complex metrics in the path integral does not lead to sensible results.
{Recently, Witten \cite{Witten:2021nzp} proposed  an admissibility criterion for complex metrics. This proposal was inspired by the work of Kontsevich and Segal  \cite{Kontsevich:2021dmb} investigating the requirements for a well-defined path integral for $p$-form fields in a complex  background metric.  See also the early work \cite{Louko:1995jw} and, for an alternative admissibility proposal, \cite{Aharony:2021zkr}.}
Since {its inception, 
Witten's} proposal  has been investigated further in various contexts (see \cite{Bondarenko:2021xvz, Lehners:2021mah, Visser:2021ucg, Loges:2022nuw, Jonas:2022uqb, Briscese:2022evf, Araujo-Regado:2022jpj} for a partial list of references related to the present work). 

In order to test the proposal, it is important to compare its predictions  to those obtained using other methods where available. One such  opportunity is provided by stationary metrics, which can be analytically continued to Euclidean signature by analytically continuing both the time coordinate some physical parameters to imaginary values (such as the angular momentum and angular potential in the case of rotating black holes). As stressed in \cite{Witten:2021nzp}, it is  not obvious that the partition function computed  using admissible complex saddles agrees with the one obtained from Euclidean saddles. 
The advantage of the  approach based on admissible complex metrics we consider in this paper is of course that it can be applied to more complicated time-dependent metrics which do not allow for an Euclidean continuation.

\begin{figure}
	\begin{center}
		\includegraphics[height=275pt]{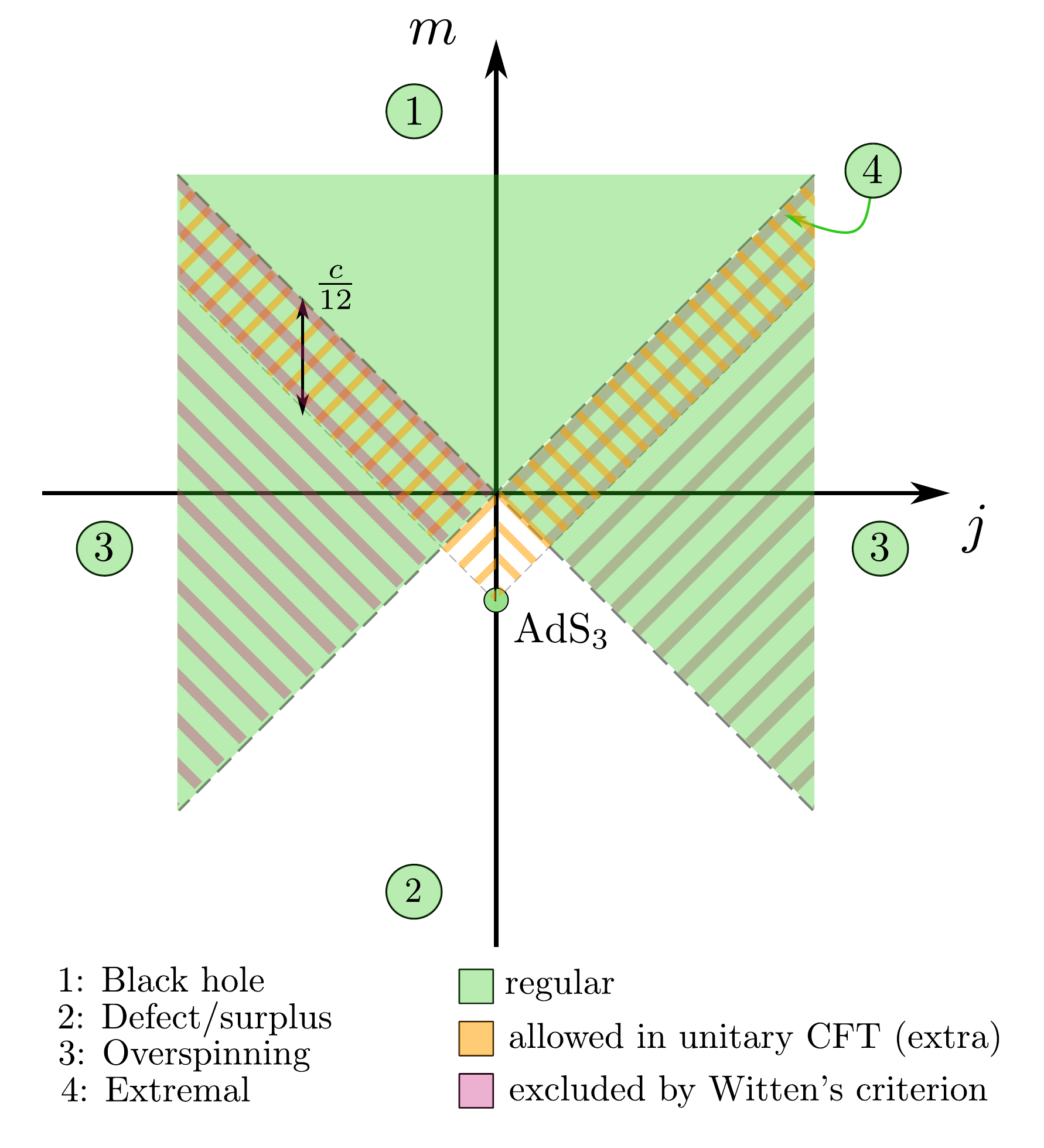}
	\end{center}\caption{A summary of our analysis. The regular geometries (in green) comprise black holes (region 1), overspinning (region 3) and pure AdS$_3$ out of the conical geometries of region 2. Of these, the admissibility criterion of \cite{Witten:2021nzp} excludes region 3, leaving only black holes and AdS$_3$. The orange region spans additional states allowed by unitarity in a dual conformal field theory.}\label{Figadm}
	\end{figure}

Anti-de Sitter gravity in $2+1$ dimensions provides a particularly rich    setting to perform such a test. One peculiarity is that, unlike in higher dimensions, the Lorentzian metrics in the overspinning regime $|J| > M$, are not nakedly singular and can be described as quotients of global AdS without fixed points\footnote{They do suffer causal pathologies such as closed timelike curves in the Lorentzian signature, but this is not a reason to exclude their  continuation to imaginary time \cite{Witten:2021nzp}.}  \cite{Hulik:2019pwr}. Furthermore,   
conical defect solutions and their spinning cousins  exist  below the  BTZ black hole threshold. Holographic conformal field theories may contain (sparse) states in both these regimes (see Figure \ref{Figadm}), making the status of these geometries as admissible saddle points  
and their potential impact on thermodynamics and the Hawking-Page transition \cite{Hawking:1982dh}
especially worth investigating. 

Motivated by these considerations, in this  note  we will consider  the family of complex metrics which arise from  analytically continuing the BTZ-type metrics \cite{Banados:1992wn} in the entire  $(M,J)$ plane to imaginary time\footnote{The geometry of the corresponding Lorentzian and Euclidean solutions  was discussed in  \cite{Banados:1992gq, Carlip:1994gc, Miskovic:2009uz, Briceno:2021dpi}.}. We study which of these `quasi-Euclidean' metrics  qualify as saddle points for the grand canonical path integral,  in the sense they are both smooth  as well as obey  Witten's admissibility criterion. The result of this analysis is illustrated in Figure \ref{Figadm}.   Our main observation is that, while smoothness alone allows for the global AdS and BTZ black hole saddles as well as the overspinning solutions,  the latter  are in fact not admissible. 
Admissibility therefore successfully excises these metrics, 
 in agreement with the Euclidean approach, where overspinning geometries do not contribute to the partition function.

 \section{BTZ class of metrics}
The BTZ metric \cite{Banados:1992wn} is the most general stationary, axisymmetric solution of the $(2+1)$-dimensional Einstein equations with negative cosmological constant, describing an object of arbitrary mass $M$ and angular momentum $J$. We will work in the
 conventions of \cite{Banados:1992gq}  and in units where the AdS radius is set to one, $l_{AdS} =1$. It is useful to define a dimensionless reduced mass and angular momentum
 \be
 m = 8 G_N M, \qquad j = 8 G_N J.
 \ee
 The BTZ metric then takes the form 
 \be 
 ds^2 = -( r^2 -  m) dt^2 +{r^2 d r^2 \over r^4 -  m r^2 + {j^2 \over 4}} +  j dt d\tilde \f + r^2 d\tilde \f^2, \label{BTZclass}
 \ee
 where $\tilde \f$ has period $2 \p$. Global AdS$_3$ corresponds to $m=-1,j=0$. We will consider the class of metrics (\ref{BTZclass}) for arbitrary real values of the parameters $m$ and $j$.
 
 For later reference, it is useful to introduce the roots $r_\pm$ of the polynomial $ x^2 - m x + j^2/4$:
 \be
r_\pm =\p \left(T_+ \pm {\rm sgn}\, (m) T_- \right),\label{defrpm}
 \ee
 where we defined\footnote{Here and in what follows, the symbol $\sqrt{\ }$ denotes the principal branch of the square root.}
 \be 
 T_\pm = {1 \over 2\p} \sqrt{m \pm j}.\label{defTpm}
 \ee
  These generically complex  parameters generalize the `left- and right-moving temperatures' of the BTZ black hole.
 The sign in (\ref{defrpm}) was chosen such that
 \be 
 {\rm Re \,} (r_+^2) \geq {\rm Re \,} (r_-^2) \label{ineqrprm}
 \ee
 for all values of $m$ and $j$.

We  rewrite the metric (\ref{BTZclass}) as
 \bea
ds^2 &=&- f dt^2 + f^{-1}dr^2 + r^2 (d  \f + N^\f dt)^2,\label{BTZmetr}\\[4pt]
f &=& r^2 - m + {j^2 \over 4 r^2}={(r^2 - r_+^2) (r^2 - r_-^2)\over r^2},\\
N^\f &=&{ j \over 2 r^2} - \O=  {r_+ r_- \over r^2}- \O.
\eea
Here, we have also made an improper coordinate transformation $ \tilde \f = \f - \O t$ which introduces an angular velocity at infinity,
\be 
ds^2\to -r^2 dt^2 + {dr^2 \over r^2} + r^2\left(d ( \f - \O t )\right)^2,
\ee
and describes AdS$_3$ in a rotating frame. Note that the Killing vector ${\pa_t}|_{\f , r}$ is timelike at infinity only if the angular velocity $\O$ satisfies 
\be
|\O| < 1.
\ee
 
It's convenient to  divide the  $m,j$ plane into 4 physically distinct  regions, see Figure \ref{Figadm}:
\begin{enumerate}
    \item  {\bf Black hole} regime, where   $m > 0,\  m > |j|$.
   The parameters  $T_\pm$  and  $r_\pm$ are real and positive, and
 $r_+$ and $r_-$ are  the locations of  the inner and  outer horizons respectively.
      \item  {\bf Defect/surplus} regime, where $ m < 0,\ |m| > |j|$.
  The parameters $T_\pm$  and  $r_\pm$ are purely imaginary. This class  includes   the global AdS metric ($m=-1, j=0$), conical defects describing backreacted point particles \cite{Deser:1983nh} ($-1< m<0,j=0$) and metrics with a conical surplus ($m<-1,j=0$), as well as spinning generalizations thereof. 
        \item {\bf Overspinning} regime, where $|m| < |j|$.
  One of the `temperatures' $T_\pm$ is real and the other is imaginary, while  the $r_\pm^2$ are complex and each others conjugate.
 \item {\bf Extremal} regime, where $m = \pm j$.
 These metrics lie on the boundaries separating regions 1, 2 and 3 and include extremal spinning black holes for $m>0$, the zero-mass limit of the BTZ black hole for $m=0$, and extremal spinning defects for $m<0$.
 \end{enumerate}

The following table summarizes the values of the parameters in these regimes:
\begin{center}
\begin{tabular}{|c |c| c| c|}
\hline
regime & $(m,j)$ & $(T_+,T_-)$ &$(r_+,r_-)$ \\ \hline \hline
1 &   $m > 0,\  m > |j|$ &  $T_\pm >0$ & $ r_+ > r_- \geq 0$\\ \hline
2 & $ m < 0,\ |m| > |j|$ & $  T_\pm  \in  i \RR $ & $r_\pm  \in  i \RR,\ 
    r_-^2 <  r_+^2\leq0$\\ \hline
3& $ |m| < |j|$&
     $ T_\pm  \in \RR, T_\mp \in i \RR$  & 
   $ r_\pm \in \CC,  r_+^2 =(r_-^2)^*$
 \\ \hline
 4 &  $m = \pm j$ & $ T_\mp =0 ,  T_\pm = {1 \over \p} \sqrt{m \over 2}$ &
 $ r_+^2 = r_-^2 = {m \over 2} $\\ \hline
\end{tabular}
\end{center}
 
It is relevant to point out that in holographic theories the unitarity bound reads
\be 
m - |j| \geq -1.
\ee
Therefore, unitary  holographic CFTs  may in principle contain states in all four of the above regimes, as illustrated in Figure \ref{Figadm}. 

\section{Quasi-Euclidean continuation}
 In the path-integral approach to quantum gravity \cite{Gibbons:1976ue}, the grand canonical partition function at inverse temperature $\b$ and angular potential $ \O, $ (with $|\O|<1$) is expressed as a path integral over metrics,
\be 
Z(\b, \O) =\tr e^{ - \b ( H + \O J)} \sim \int [Dg] e^{ i S[g]}\label{Zgc},
\ee
where the metrics should behave near infinity as
\be \label{asqE}
ds^2 \to r^2 d\t^2 + {dr^2 \over r^2} + r^2\left(d \f - i  \O d\t \right)^2
\ee
with coordinates having the periods
\be
(\t ,\f ) \sim (\t +\b , \f ) \sim (\t, \f + 2 \p).
\ee
The asymptotic condition shows that the metrics included in the measure should be allowed to be complex. However, including  all complex metrics\footnote{We define a complex metric to be a complex, invertible, symmetric (0,2) tensor.}  does not lead to sensible results. Witten's admissibility criterion \cite{Witten:2021nzp}, which we review in Section~\ref{sec:admissability}, is a proposal to restrict  to a subclass of physically sensible complex metrics.

In the semiclassical limit $G_N \to 0$, the path integral will be dominated by classical saddle points. 
A large class of potential classical saddles satisfying (\ref{asqE}) is obtained from the metric (\ref{BTZmetr}) by continuing the time coordinate to  the imaginary axis,
$t \to i \t$. The resulting metric is said to be quasi-Euclidean (qE) and takes the form 
\be
ds^2_{qE} = f d\t^2 + f^{-1}dr^2 + r^2 \left(d \f +i N^\f d\t \right)^2.\label{BTZcompl1}
\ee
Note that, as emphasized in the Introduction, in this approach one does  {\em not} continue the parameters $j$ and $\O$ to imaginary values, which would result in a real  Euclidean metric.

The metrics (\ref{BTZcompl1}) generically have coordinate singularities where the metric degenerates. As in (pseudo-)Riemannian geometry, these can be
either an artifact of the coordinate system (and disappear upon making a suitable coordinate change), or reflect a genuine pathology. 
We will interpret the latter case as a sign that the classical gravity approximation breaks down and we should not include the solution as  a  saddle.

Our strategy will be to first determine  the  qE metrics which are smooth and subsequently, in Section~\ref{sec:admissability}, analyze  which ones obey Witten's admissibility criterion.

\section{Smoothness}
In order to  analyze the smoothness of the quasi-Euclidean metrics, it is useful to make a further coordinate redefinition.
In the above coordinate system, the quasi-Euclidean (qE) metric (\ref{BTZcompl1}) degenerates at $r=0$, where $\det g$ vanishes.
Unless $r_+$ or $r_-$ also vanishes, this is a coordinate singularity which can be removed by defining the new coordinate
\be 
u = r^2.
\ee
The qE metric then becomes
\be
ds^2_{qE} ={ (u-r_+^2)(u-r_-^2)\over u} d\t^2 + {du^2 \over 4 (u-r_+^2)(u-r_-^2)}  + u \left(d \f +i \left( {r_+ r_- \over u} -\O\right) d\t \right)^2.\\
\label{BTZcompl}
\ee
Coordinate singularities still occur when $u=r_+^2$ and $u=r_-^2$, while $\det g$ is constant for all values of $u$. This of course happens only if $r_+^2$ and $r_-^2$ are real, i.e. in the regimes 1, 2 and 4, while the overspinning regime 3 is free of coordinate singularities. When $r_+^2$ and $r_-^2$ are real it will be useful to divide the spacetime into subregions to the left (L), in the middle (M) and to the right (R) of the coordinate singularities\footnote{We note that the qE metric is invariant under the formal involution
$
u \to - u,  
\t \to \pm i \t , 
\f \to \pm i \f , 
r_\pm \to \pm i r_\pm, 
$
where the sign is fixed by requiring the inequality (\ref{ineqrprm}) to hold. This involution exchanges the regions as follows:
$
1L \leftrightarrow  2 R,  
1M  \leftrightarrow  2 M, 
1R  \leftrightarrow 2 L .
$
Some of our results below can be seen as consequences of this property.
}:
\bea
L&:&\qquad u \leq r_-^2,\\
M&:&\qquad r_-^2 < u < r_+^2,\\
R&:&\qquad  r_+^2 \leq  u.
\eea
We note that in the extremal regime 4 the middle (M) region is absent.

In the rest of this section we want to establish in which regions and for which values of the parameters $\b, \O$ the qE metrics (\ref{BTZcompl}) are invertible and smooth.
 In three-dimensional gravity, all curvature invariants are locally constant and detecting singularities is more subtle  than in higher dimensions, requiring a careful description of the global manifold structure.
Extending the Lorentzian analysis \cite{Banados:1992gq, Cangemi:1992my}, we will use the fact that the `quasi-Euclidean manifold'  on which the qE metric is defined can be obtained from the smooth group manifold $SL(2,\CC)$ by two operations:  a quotient  by a discrete group and a subsequent restriction to a three-dimensional real subspace. In sections \ref{Secquotient} and \ref{Secreal}  we establish when these operations introduce singularities. A complementary 
approach, presented in  Section \ref{Sechols}, is to study the holonomy of the Chern-Simons connections, which for regular solutions should be trivial when evaluated on smoothly contractible cycles. This will lead to results consistent with the first approach, though is somewhat more crude since, as we shall see,  certain singularities in the manifold structure can still lead to trivial Chern-Simons holonomy.

\subsection{The quotient}\label{Secquotient}
The $SL(2,\CC)$ group manifold can be viewed as  a smooth hypersurface in $\CC^4$,
\be 
\det G = G_{11} G_{22} - G_{12}G_{21} = 1,\label{H}
\ee
with invariant metric
\be 
ds^2 = - \half\, \tr \left(d G^{-1} dG\right) = -  d G_{11} dG_{22} + dG_{12} dG_{21}.\label{invsl2}
\ee
When $m>0$, the aforementioned  quotient amounts to 
imposing the following two identifications:
\bea 
I_1: \qquad (G_{11} , G_{22} , G_{12}, G_{21}) &\sim &   (e^{ 2 \p r_+ } G_{11} , e^{-2 \p r_+ } G_{22} , e^{- 2 \p r_- }G_{12}, e^{ 2 \p r_- }G_{21})\nonu
I_2: \qquad (G_{11} , G_{22} , G_{12}, G_{21}) &\sim & (e^{  i (r_- - \O r_+) \b } G_{11} ,e^{-  i (r_- - \O r_+) \b } G_{22} , \nonu
 && e^{-  i (r_+ - \O r_-) \b }G_{12},  e^{  i (r_+ - \O r_-) \b } G_{21}),\label{Is}
\eea 
while for $m<0$, one has to interchange  $r_+$ and $r_-$ in these expressions. Note that these identifications preserve the hypersurface \eqref{H}. Their appropriateness will become manifest in Section~\ref{Secreal}, where we will recover the qE metric \eqref{BTZcompl} by selecting a real slice of a group element satisfying \eqref{Is}. We should also note that (\ref{Is}) breaks down in  the limit of extremal metrics, i.e.\ regime 4,  the details of which are  discussed in Appendix \ref{Appextr}.  

The quotient space is a smooth manifold only if the identifications act without fixed points on the hypersurface (\ref{H}), see e.g.\ \cite{HawkingEllis}. 
Both identifications have a fixed point in $\CC^4$ at $G_{11}= G_{22}=G_{12}=G_{21}=0$, which however does not lie on  (\ref{H}).
From the form of (\ref{Is}) we see that singularities can occur   on two possible loci, namely 
\begin{itemize}
\item at $G_{12}= G_{21} = 0$, $G_{11}G_{22}=1$. This is a fixed locus of $I_1$ for $r_+ =0$ provided that $r_- \neq i$. The case  $r_+ =0$, $r_- = i$, is special, since $I_1$ then becomes trivial. Similarly, it is a fixed locus of $I_2$ for $\O = {r_-\over r_+}$ provided that $\b \neq  { 2 \p r_+ \over r_+^2 - r_-^2}$, with the case $\O = {r_-\over r_+}$, $\b =  {2 \p r_+ \over r_+^2 - r_-^2}$ corresponding to $I_2$ becoming trivial.
\item at $G_{11}= G_{22} = 0$,  $G_{12}G_{21} = -1$. This is a fixed locus of $I_1$ for $r_- =0$ and $r_+ \neq i$, and  a fixed locus of $I_2$ for $\O = {r_-\over r_+}$ and $\b \neq  {2 \p r_+ \over r_+^2 - r_-^2}$.
\end{itemize}
As one would expect, these loci correspond to the coordinate singularities at $u=r_+ ^2$  or $u=r_- ^2$ as we shall see shortly.
Recalling the ranges of the parameters $r_\pm$ in  the various regimes   and the fact that $\b$ and  $\O$ are required to be real, the above analysis (and its extension to region 4 in Appendix \ref{Appextr}) can be summarized in the following table of regular quotients:
\begin{table}[!h]
\begin{center}
\begin{tabular}{|c |c| }
\hline
regime &  quotient regular for\\ \hline \hline
1  & $\left\{\begin{array}{l} \O = {r_- \over r_+}, \b = { 2 \p r_+ \over r_+^2 - r_-^2}\\
\O = {r_+ \over r_-}, \b = { 2 \p r_- \over r_+^2 - r_-^2}\\
\O \neq \{ {r_- \over r_+}, {r_+ \over r_-} \}\end{array}\right. $\\ \hline 
2  & $\left\{\begin{array}{l} r_- = i, r_+ = 0\\
\O \neq \{ {r_- \over r_+}, {r_+ \over r_-} \}\end{array}\right. $\\ \hline 
3 & always \\ \hline  
4  & $\left\{\begin{array}{l} \O = {\rm sgn}\, {m \over j}, \b = \infty , m > 0\\
\O \neq  {\rm sgn}\,  {m \over j} , m \neq 0 \end{array}\right. $\\ \hline 
\end{tabular}
\end{center}
\caption{Summary of smooth quotients.\label{Tablesmoothquotients}}
\end{table}\\
A remark  is in order concerning this table. In principle, we could have also allowed  the values of $\b$ in the first and second lines and of $r_-$ in the fourth line to be an integer multiple of the displayed values, since the relevant identification would then still act trivially. However, in doing so we would no longer describe a quotient of $SL(2, \CC)$ but rather a multi-sheeted covering space. Furthermore, the resulting manifold would not be a smooth covering space (see e.g.\ \cite{Lee}), since the sheets would meet at the   loci described above, and therefore we will not consider these spacetimes in this context\footnote{Such singularities do have an interesting holographic interpretation as arising from insertions of degenerate primaries, see \cite{Castro:2011iw, Perlmutter:2012ds, Campoleoni:2013iha, Raeymaekers:2014kea}.}.   

\subsection{The real slice}\label{Secreal}
Now we turn to the second operation of restricting the above quotients to a real, three-dimensional subspace. For this purpose we parametrize the group element as
\bea
G_{11} &=&   z_1 w \exp \left({   r_+ \f}+     i (r_- - \O r_+) \t\right)\nonu
G_{22} &=&   z_1 w^{-1} \exp \left[- \left({  r_+ \f}+     i (r_- - \O r_+) \t \right)\right]\nonu
G_{12} &=&   z_2 w^{-1}  \exp \left[ - \left({  r_- \f}+      i (r_+ - \O r_-) \t \right)\right]\nonu
G_{21} &=&   z_2 w \exp \left( { r_- \f}+    i (r_+ - \O r_-) \t\right),
\label{coords1}
\eea 
where $z_1, z_2 \in \CC, w \in \CC \backslash \{ 0 \}$ and $\f,\t \in \RR$.
For instance, in regime 1 one checks that these are good coordinates away from the locus where at least one of the $G_{ij}$ vanishes, as long as $r_+ \neq r_-$ and $\O \neq -1$.  The equation (\ref{H}) becomes
\be 
z_1^2 - z_2^2 =1,
\ee
while the identifications (\ref{Is}) read
\be 
I_1: \f \sim \f + 2 \p, \qquad I_2: \t \sim \t+ \b .
\ee

\subsubsection*{Regime 1}
Let us discuss the real slice in terms of these coordinates firstly in regime 1. The qE geometry in the subregions 1R, 1M, and 1L arises from  imposing:
\begin{itemize}
    \item Region 1R:
$ 
w= 1, \arg z_1 = \arg z_2 = 0 .
$
  \item Region 1M:
$ 
w= 1, \arg z_1 = 0, \arg  z_2 = {\p \over 2} .
$
 \item Region 1L:
$ 
w= 1, \arg z_1=  \arg  z_2 = {\p \over 2} .
$
\end{itemize}
In all cases, $z_1$ and $z_2$ can be expressed in terms of the real coordinate $u$ as
\be 
 z_1 = \sqrt{ u - r_-^2 \over r_+^2- r_-^2},  \qquad z_2 = \sqrt{ u - r_+^2 \over r_+^2- r_-^2}.
 \ee
 One checks that the pullback of the invariant metric  (\ref{invsl2}) is indeed (\ref{BTZcompl}).
 
 Let us now discuss in which cases the pullback  to the real slice of the smooth metric on the  regular quotient spaces listed in Table \ref{Tablesmoothquotients}  
 fails to be a smooth complex metric.
 For $\O = {r_-\over r_+}$ and $\b = {2\p r_+ \over r_+^2 - r_-^2}$, 
 the locus $u \to r_+^2$ has codimension 2 and is the $\f$-circle embedded as
 $ G_{12}=G_{21}=0, G_{11} = G_{22}^{-1}= e^{ r_+ \f}$. In other words, the $\t$-circle `pinches off' there. From the form of $G_{12},  G_{21}$ we see that it does so smoothly and that the  spacetime  locally looks like $\RR^2 \times S^1$. In this case the region 1R  ends smoothly at $u \to r_+^2$. Similarly, for  $\O = {r_+\over r_-}$ and $\b = {2\p r_- \over r_+^2 - r_-^2}$ the region 1L forms a smooth submanifold.
 
 For $\O \neq {r_-\over r_+}$, the locus $u =  r_+^2$ has codimension 1: it is an $S^1 \times S^1$ embedded as \be G_{12}=G_{21}=0, \qquad G_{11} = G_{22}^{-1}= e^{ r_+ \f +  2 \p  i (r_- - \O r_+) \t}.\label{embhor}\ee 
 Neither the $\t$ or $\f$-circles pinch off and one expects the spacetime  to continue into region 1M.  However, 
 the regions 1R and 1M cannot be joined together smoothly\footnote{Independently of this, we will see in Section \ref{sec:admissability}  that admissibility would discard these joined spacetimes.} 
 as we shall now argue in two complimentary ways: the first based on degeneracy of the pulled-back metric and the second from the fact that the embedded  real submanifold is not sufficiently smooth. 
 
 Firstly, one can argue that the coordinate singularity in $u = r_+^2$ is not removable. Indeed, from (\ref{embhor}) we see that  $\pa_\f$ and
 $\pa_\t$ provide linearly independent basis vectors of the tangent space at each point. The metric on this locus is however always of rank one, independent of the chosen  coordinates on the real slice. Indeed, the metric (\ref{invsl2}) pulled back to the $u = r_+^2$ locus $ G_{12}=G_{21}=0, G_{11} = G_{22}^{-1}$ is 
 \be 
 ds^2_{hor} =  { dG_{11}^2 \over G_{11}} .
 \ee
  Since a nondegenerate metric should have maximal rank when pulled back to a proper submanifold, we see that the quasi-Euclidean `metric' is in this case degenerate.
 
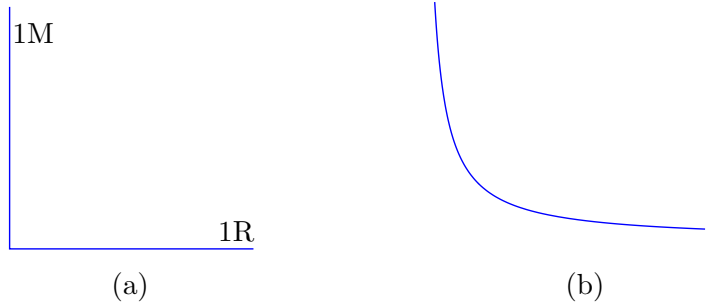
\begin{figure}
 \begin{picture}(300,100)
	\put(100,10){\includegraphics[height=95pt]{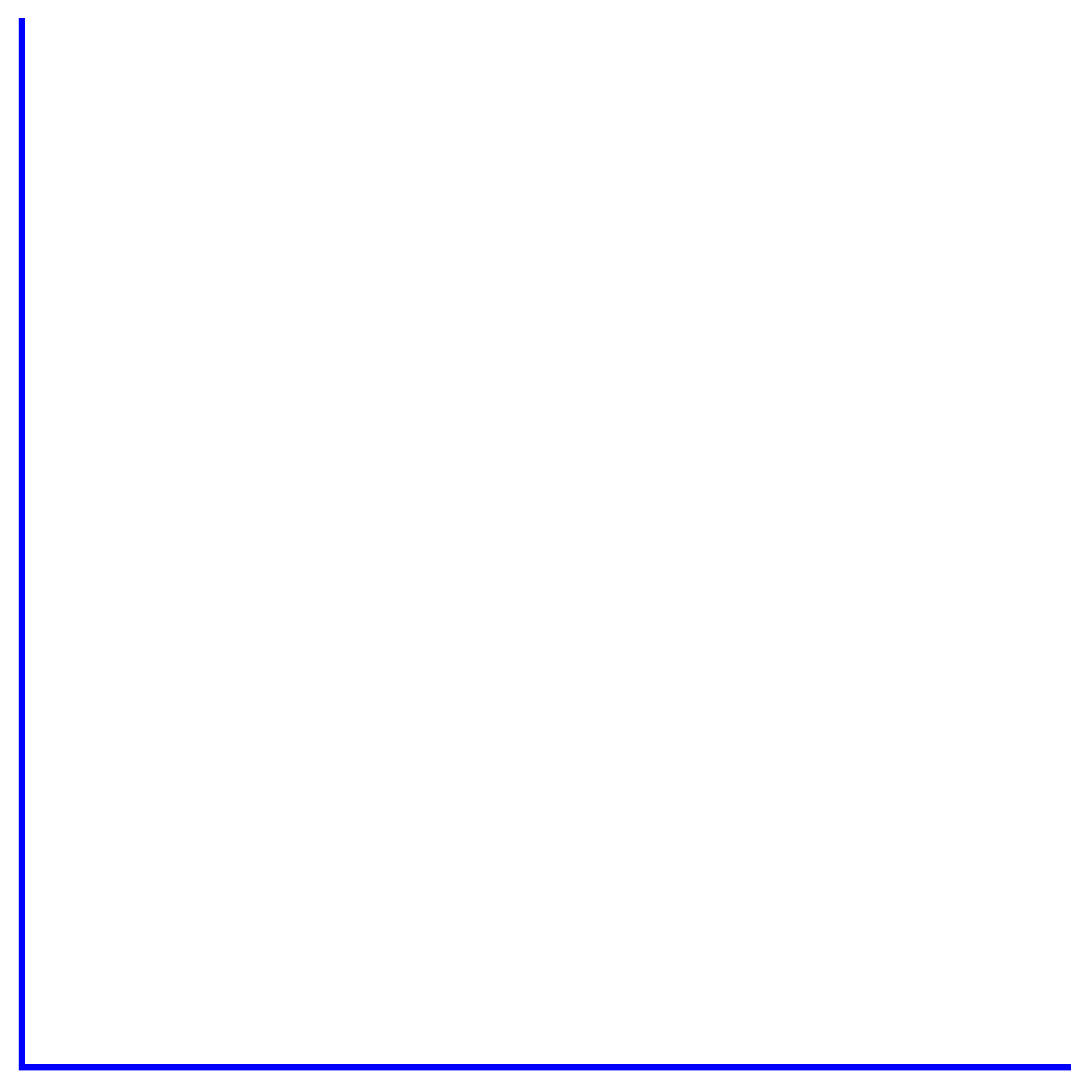}}
	\put(250,10){\includegraphics[height=100pt]{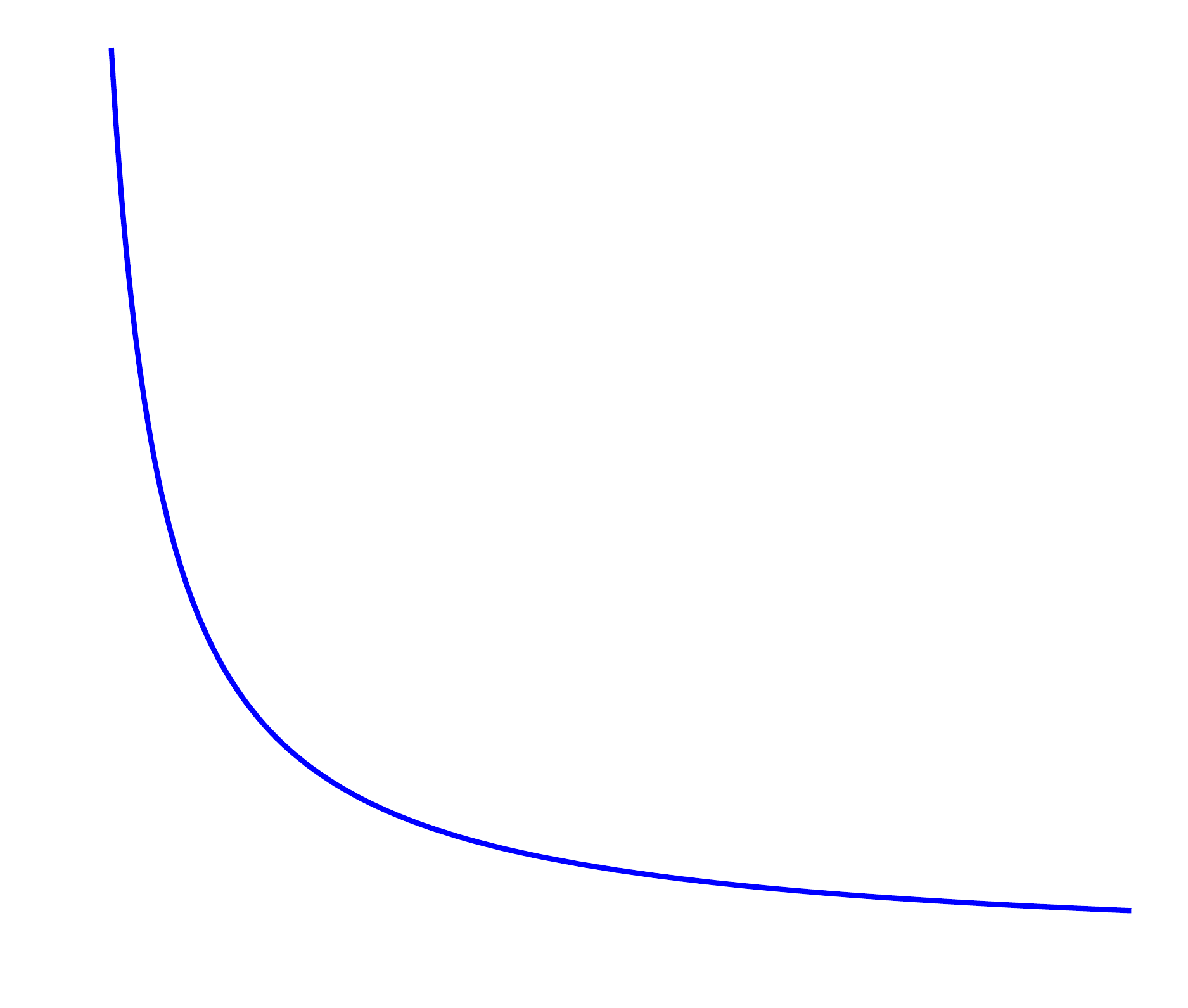}}
	\put(140,-5){(a)}
	\put(180,15){1R}
	\put(103,90){1M}
	\put(310,-5){(b)}
	\end{picture}
	\caption{Radial coordinate lines embedded in the $G_{12}$ plane in (a) Region 1 for $\O \neq {r_- \over r_+}$ (b) Region 3.}\label{Figreal}
	\end{figure}

 A second argument comes from considering the smoothness  of the 3D submanifold.
  It is instructive to see how the radial coordinate lines of constant $\f$ and $\t$ are embedded in the ambient space in the vicinity of $u = r_+^2$.
 While they are embedded  as straight lines in the $G_{11}$ and $G_{22}$ planes, they are not  embedded  as  differentiable curves in the $G_{12}$ and $G_{21}$ planes, where they have a rectangular `corner' as shown in Figure \ref{Figreal}(a).

 Therefore, the radial coordinate lines and the embedded 3D real manifold   are  at most of class $\calc^0$  (see \cite{HawkingEllis}). 
However,  in order to be able to reliably compute the curvature of the pulled-back metric, one would need the   embedding map to be at least of class $\calc^2$.   
Similarly, one argues that, for $\O \neq {r_+\over r_-}$, the   region 1L cannot be not smoothly  joined to the region 1M at $u = r_-^2$.

\subsubsection*{Regime 2}
In region 2, the real slice in subregions 2R, 2M, and 2L is:
\begin{itemize}
    \item Region 2R:
$ 
w= 1, \arg z_1 = \arg z_2 = 0 .
$
  \item Region 2M:
$ 
w= 1, \arg z_1 = {\p \over 2} , \arg  z_2 = 0 .
$
 \item Region 2L:
$ 
w= 1, \arg z_1=  \arg  z_2 = {\p \over 2} .
$
\end{itemize}
In all cases, $z_1$ and $z_2$ can be expressed in terms of the real coordinate $u$ as
\be 
 z_1 = \sqrt{ u - r_+^2 \over r_+^2- r_-^2},  \qquad z_2 = \sqrt{ u - r_-^2 \over r_+^2- r_-^2}.
 \ee
Similarly as in  regime 1, one argues that for $r_- = i, r_+ =0$, the region $2R$ is a smooth submanifold which ends at $u=0$, where the $\f$-circle pinches off smoothly.
Also, as above one shows that for  $r_- = i, r_+ =0$, the regions 2L and 2M cannot join smoothly.   Similarly, when $\O \neq   {r_- \over r_+}, r_+ \neq 0$,  the region 2R cannot be smoothly joined to the region 2M.

\subsubsection*{Regime  3}
 In the overspinning regime, the real slice is defined by setting 
 \be
 w=1, \qquad z_1 = \sqrt{ u - r_-^2 \over r_+^2- r_-^2},  \qquad z_2 = \sqrt{ u - r_+^2 \over r_+^2- r_-^2}.\label{embedregion3}
 \ee
As already mentioned, the metric is everywhere nondegenerate and one checks that (\ref{embedregion3}) defines a smooth submanifold; for example, the radial curves are now smooth curves in the ambient $\CC^4$ as illustrated in Figure \ref{Figreal}(b).

\subsubsection*{Regime  4}
Using the embedding (\ref{Gextremal}) in  Appendix \ref{Appextr}, one can similarly show that the real slice defining the extremal metrics is not smooth  for $\O \neq  {\rm sgn}\,  {m \over j}$, $m \neq 0$.  Smooth extremal metrics can be thought of  as $j \to \pm m$   limits   of the smooth metrics in region 1, in the sense of requiring $\b \to \infty, \O \to \pm 1$ while keeping $\b (1-\O^2)= {4 \p \over m} $ fixed.

Combining the above results, we arrive at Table \ref{Tablesmooth} of smooth qE metrics. We also indicate in the last column whether the spacetime is a potential saddle for the partition function $Z(\b, \O)$ in (\ref{Zgc}). For this it needs to contain the region $u \to \infty$ and  satisfy $|\O| < 1$.
\begin{table}[!h]
\begin{center}
\begin{tabular}{|c ||c| c|}
\hline
region & regular if  &potential saddle?\\
\hline \hline
 1R & $\b = {2 \p r_+\over r_+^2 - r_-^2}, \O = {r_- \over r_+} $ & $\times$ \\
\hline
 1L & $r_- \neq 0, \b = {2 \p r_-\over r_+^2 - r_-^2}, \O = {r_+ \over r_-} $ & -\\ 
\hline \hline
  2R &  $ r_+= 0, r_- = i, \b \in \RR ,  $ & $\times$\\
\hline\hline
3 & $\b, \O \in \RR$ & $\times$\\
\hline \hline
4L  & $\b \to  \infty, \O \to {\rm sgn} {j \over m}$ & - \\
\hline 
4R  & $\b \to \infty, \O \to {\rm sgn} {j \over m}$ & $\times$\\
\hline  
\end{tabular}
\end{center}
\caption{Summary of smooth qE metrics.} \label{Tablesmooth}
\end{table}

\subsection{Smoothness and holonomies}\label{Sechols}

In order to complement the above discussion on regularity, one can study holonomies of the Chern-Simons connections associated to the metric \cite{Cangemi:1992my}. These quantities encode information on the topology of spacetime, but they can still fail to detect certain singularities. Despite these shortcomings, one can at least compute the relevant holonomies as a consistency check.

The flat $SL(2, \CC) \times SL(2, \CC)$ gauge potentials describing the quasi-Euclidean metrics (\ref{BTZcompl}) are of the form
\be 
A = g^{-1} dg, \qquad \tilde A = \tilde g^{-1} d\tilde g.
\ee
The  $SL(2, \CC) $ group elements can be 
can be obtained from the group element $G$ by writing it as
\be
G = g \tilde g^{-1}, \qquad g = e^{ \p T_- \left(\f - i (1+ \O ) \t\right)\s_3 } B(u), \qquad 
\tilde g = e^{- \p T_+ \left(\f + i (1- \O ) \t\right)\s_3} B(u)^{-1},\label{groupelsCS}
\ee
where the specific form of the $2\times 2$ matrix $B(u)$ is not needed, since we shall seek trivial holonomies around curves with constant $u_0$.

Using (\ref{groupelsCS}) one can easily compute the holonomies of $A$ and $\tilde A$ around closed curves $\g$ at constant radius $u = u_0$. For an angular circle of period $2\pi$ at constant $\tau = \tau_0$ one finds
\bea
H_\g &=& B(u_0)^{-1}  e^{ 2 \p^2 T_- \s_3} B(u_0) \, \\
\tilde H_\g &=& B(u_0)  e^{- 2 \p^2 T_+ \s_3} B(u_0)^{-1} \, .
\eea
These are trivial (meaning equal to $1$ or $-1$, i.e.\ in the center of $SL(2,\CC)$) for $r_\pm \in i \mathbb{Z}$, namely for
\be 
m = - { p^2 + q^2\over 2}, \qquad j = - { p^2 - q^2\over 2}, \label{eq:trivial_CS_holonomy}
\ee
where $p,q$ are nonzero integers. In the extremal case, when either of them vanishes, one needs to use a different group element. We discuss this subtlety in Appendix~\ref{Appextr}.
Except for the pure AdS$_3$ case $p=q=1$, these represent a discrete family of conical surpluses and their spinning generalizations. Such branched covering spaces are singular as manifolds as we have explained below Table \ref{Tablesmoothquotients}.
In region 1 (black hole regime) the angular circle is not contractible, so there is no issue.

For a time circle with period $\b$ at constant $\phi = \phi_0$, the holonomies read
\bea
H_\g &=& B(u_0)^{-1}  e^{-i \p T_- (1+\O) \b \s_3} B(u_0) \, ,\\
\tilde H_\g &=& B(u_0)  e^{-i \p T_+ (1-\O) \b \s_3} B(u_0)^{-1} \, ,
\eea
which are in the center only for 
\bea
\b &=& \pi \, \frac{p(r_+ + r_-) + q(r_+ - r_-)}{r_+^2-r_-^2} \, , \\
\O &=& \frac{p(r_+ + r_-) - q(r_+ - r_-)}{p(r_+ + r_-) + q(r_+ - r_-)}
\eea
with $p \, , \, q \in \mathbb{Z}$. However, most of these solutions again describe singular branched covering spaces. The inequivalent ``minimal'' choices are (representable by) $p=q=1$ and $p=-q=1$. In region 1 the former is consistent with the regularity conditions of the black hole geometry (region 1R), while the latter gives region 1L which is not a potential saddle for the partition function.

In region 2, these solutions would lead to imaginary $\beta$ which gives a singular geometry. A similar argument excludes region 3. All in all, the regular geometries that we have determined do have trivial Chern-Simons holonomies along contractible curves, as expected, but this condition by itself is not sufficient.

\section{Admissibility }\label{sec:admissability}
In this section we investigate in which regions the qE metrics (\ref{BTZcompl})  obey Witten's admissibility  criterion. {As discussed in \cite{Kontsevich:2021dmb, Witten:2021nzp}, defining a well-behaved (semiclassical) path integral on a manifold requires that large field fluctuations be suppressed by the (exponential of the) Euclidean action. For complex metrics coupled to $p$-form fields, the real part of the action ought to be positive definite, or more generally bounded from below. Extending this criterion to fields of different type is more subtle, due to the difficulties in coupling them consistently to gravity.}

In 3 dimensions, the admissibility conditions   reduce   to \cite{Witten:2021nzp}
\be 
{\rm Re\,} (\sqrt{g}) >0, \qquad {\rm Re \,} (\sqrt{g} \l_i^{ -1}) >0,
\ee
where $\l_i, i = 1,2,3$ are the eigenvalues of the metric. Applying to (\ref{BTZcompl}), the first condition is satisfied since $\sqrt{g} = \half$.
The remaining conditions reduce to the requirement that $g_{uu}>0$ and  that the eigenvalues of the 2D $\t, \f$ submatrix have positive real parts. As shown in \cite{Witten:2021nzp}, this  latter condition is equivalent to $g_{\t\t} >0$. Using (\ref{BTZcompl}), the admissibility criteria $g_{uu}>0$, $g_{\t\t} >0$ then reduce to\footnote{We note  that, for positive $u$, the second inequality implies the first.}
\begin{align}
    I: & &(u - r_+^2)(u-r_-^2) &>0,\\
    II: & &(1- \O^2)u - (r_+^2 + r_-^2) + 2 r_+ r_- \O &>0.
\end{align}
The second condition, which is linear in $u$, is always satisfied on a half-infinite line in the $u$ coordinate and can be rephrased as
\begin{align}
 II: & &
\left\{ \begin{array}{ll}
      u > u_0    &{\rm for\  } |\O|<1\\
     ( r_+ \mp r_-)^2 <0    &{\rm for\  } \O = \pm 1\\
       u <  u_0    &{\rm for\  } |\O|>1
     \end{array}\right. ,
\end{align}
where $u_0$ is the zero of $g_{\t\t}$, namely
\be 
u_0 = {r_+^2 + r_-^2 - 2 r_+ r_- \O \over 1- \O^2}.
\ee

We now investigate in which of the  regions  described above, i.e. (1L, 1M, 1R; 2L, 2M, 2R; 3; 4L, 4R), the conditions $I$ and $II$ are satisfied, possibly upon imposing some restriction on the angular velocity  $\O$. In regions 1M, 2M the condition $I$ is violated. In region 1L one finds that $I$ and $II$  hold simultaneously only if $u_0 = r_-^2 $ and $|\O|<1$, while in region 1R they hold if $u_0 = r_+^2 $ and $|\O|>1$. In region 2L, conditions $I$ and $II$ both hold for any $|\O | >1$, while in region  2R they hold for $|\O | <1$.  In region 3, condition $II$ is always violated in some range of $u$ (namely $u<u_0$ when $|\O|<1$ and $u>u_0$ when $|\O|>1$). In region 4, taking $m = j$ for definiteness, condition I is obeyed (except at $u = m/2$). When $\O <1$,  condition II is violated  in region 4R for $ {m \over 2} < u < {m \over \O + 1}$, while for $\O >1$,  it is violated  in region 4L for $  {m \over \O + 1} < u < {m \over 2} $. Thefore both 4L and 4R become admissible in the limit $\O \to 1$.  These conclusions are  summarized in the following table:
\begin{table}[!h]
\begin{center}
\begin{tabular}{|c |c|}
\hline
region & 
 admissible if \\ \hline  \hline
1L& $\O = {r_+ \over r_-}$ \\
1M& never\\
1R& $\O = {r_- \over r_+}$ \\\hline
2L& $|\O|>1$\\
2M& never \\
2R& $ |\O|<1$ \\\hline
3 & never\\ \hline
4L & $\O \to {\rm sgn}\, {j \over m}$\\
4R & $\O \to {\rm sgn}\, {j \over m}$\\ \hline
 \end{tabular}
\end{center}
\caption{Summary of admissible qE metrics.\label{Tableadm}}
\end{table}

\section{Discussion}
Combining the results from the smoothness (Table \ref{Tablesmooth}) and admissibility (Table \ref{Tableadm}) analyses, we conclude that the smooth, admissible quasi-Euclidean saddles contributing to the partition function (\ref{Zgc}) are
\begin{center}
\begin{tabular}{|c |c| }
\hline
region & 
 smooth \& admissible for \\ \hline  \hline
 1R & $\b = {2 \p r_+\over r_+^2 - r_-^2}, \O = {r_- \over r_+} $\\ \hline
 2R &  $ r_+= 0, r_- = i, \b \in \RR, |\O|<1 ,  $\\
 \hline
 \end{tabular}
\end{center}
In other words, the contributing saddles are the black holes with the standard relations between 
$\b, \O$ and $m,j$, and rotating thermal  AdS.
Therefore the admissible complex saddles agree\footnote{Similar conclusions were reached, from a quite different approach, in \cite{Afshar:2017okz}.} with those considered in the more standard approach of  going to Euclidean signature\footnote{In this approach, one finds an additional $SL(2,\ZZ)$ family of Euclidean instantons, which reproduce the modular properties of a  dual Euclidean CFT. However, upon continuing back the angular potential $\O_{Eucl} \to i \O$, these would have complex action, making their relevance for  real-time thermodynamics unclear. } by continuing also the angular momentum and the angular potential  to imaginary values  \cite{Maldacena:1998bw}.   For this agreement it was crucial that admissibility discards the overspinning metrics. These would be hard to interpret thermodynamically as they are wormhole-like geometries connecting two asymptotic regions, and do not contribute in the Euclidean approach.
We see it as an encouraging sign that  the method passes this nontrivial consistency check. 

The computation of the contribution of these saddles to the partition function (\ref{Zgc}) proceeds in the standard manner using holographic renormalization \cite{Henningson:1998gx,Balasubramanian:1999re}.
The regularized on-shell action is 
\be
S_{reg} =  {1 \over 4\p G_N}\left[ \int_{u\leq L^2} d^3 x \sqrt{-g}  -\half  \int_{u =L^2}d^2 x \sqrt{-\g} (K-1)\right],\label{Sreg}
\ee
where $L$ is a large radius cutoff, to be taken to infinity in the end. 
Evaluating on a solution,  the divergent terms of order $L^2$ are cancelled 
and taking $L$ to infinity the result is
\be
i S_{ren} =  { c \over 12} \b (2 u_0 -m  ).\label{renaction}
\ee
Here, $c = {3 \over 2 G_N}$ and $u_0$ is the starting point of the  radial interval $u_0 \leq u\leq L^2$.
For both saddles, the starting point is at $u_0 = r_+^2$, leading to 
\bea 
Z (\b, \O)  &=& e^{i S_{ren,1R} } +  e^{i S_{ren,2R} } = e^{{  c \over 3}{ \p^2 \over \b( 1 - \O^2)} } +  e^{{  c \over 12} \b }.
\eea
The Hawking-Page transition \cite{Hawking:1982dh} arises from exchange of dominance between these saddles in $(\b, \O )$ space (see also \cite{Kurita:2004yn}).

\section*{Acknowledgements}
The research of JR was supported
by the Grant Agency of the Czech Republic under the grant EXPRO 20-25775X. The research of AC and IB was partially supported by the Fonds de la Recherche Scientifique - FNRS under Grants F.4503.20 and T.0022.19.
The authors gratefully acknowledge bilateral travel support from the Mobility Plus Project FNRS 20-02 and the PINT-BILAT-M grant R.M005.19.

\begin{appendix}

\section{Details of the extremal case}\label{Appextr}
Now let us consider the extremal metrics where $m = \pm j$. Without loss of generality (by making a parity transformation if necessary), we can assume that
\be 
m = j
\ee
so that \be T_- =0, \qquad T_+ = {r_+ \over \p} = {1 \over \p} \sqrt{m\over 2}.
\ee

In this case, it turns out that the identifications act on the complex $2 \times 2$ group element $G$ as follows:
\bea 
I_1: \qquad G &\sim &e^{ \sqrt{2} \p \s_+} G  e^{ \sqrt{2} \p (\s_- + m \s_+)} ,\\
I_2: \qquad G &\sim &e^{- {i \over  \sqrt{2}} (1+\O )\b \s_+} G  e^{ {i \over  \sqrt{2}} (1-\O )\b (\s_- + m \s_+)},
\eea
where $\s_\pm = \half (\s_1 \pm i \s_2)$.
Aside from $G=0$, these have the following fixed points. $I_1$ has  fixed points only when $m=0$, namely at $G_{21}=-G_{12}, G_{22}=0$. The identification $I_2$ has fixed loci for $\O = 1$, namely at $G_{21} = G_{22} =0$ and for $m=0$ at $G_{21} = {{1 + \O\over 1-\O}} G_{21}$. This leads to the smooth quotient spaces given in Table \ref{Tablesmoothquotients}.

The real slice is in this case defined by taking the group element to be of the form
\be 
G= \exp {\left( \f - i (1 + \O) \t\right) {\s_+ }\over \sqrt{2} } \left( \begin{array}{cc}{1 \over \sqrt{a}}&0\\0 & \sqrt{a} \end{array} \right)
\exp{\left( \f + i (1 - \O) \t\right) \left(\s_- + m\s_+ \right)\over \sqrt{2}},\label{Gextremal}
\ee
where
\be 
a= 2 u - m  = 2 (u - r_+^2).
\ee 
One sees from (\ref{Gextremal}) that in this case the horizon $u \to r_+^2$ corresponds points at infinity in $SL(2,\CC)$:
\be 
G_{11} \to \infty,\quad  G_{22}\to 0,\quad G_{12} \to \infty, \quad G_{21}\to 0 .
\ee

\end{appendix}

\bibliographystyle{ytphys}
\bibliography{allowable}

\end{document}